# Perturbation-based Compensation with EEPN-free Phase Recovery as Back Propagation

Chuang Xu(1), Alan Pak Tao Lau(1)

(1) Photonics Research Institute, Department of Electrical and Electronic Engineering, The Hong Kong Polytechnic University, Hong Kong SAR, China. chuang.xu@connect.polyu.hk

**Abstract** *We propose a feed-forward perturbation-based method that uses the noisy received signal to compensate for nonlinear distortion, which outperforms the conventional decision-based method and avoids decision feedback. Additionally, combining it with the EEPN-free carrier phase recovery shows additional gain due to a fully symmetrical propagation-backpropagation structure.* 

## Introduction

Perturbation-based (PB) analysis provides analytical approximations for fiber nonlinearity (NL)-induced distortion, facilitating the design of pre-distortion (PD) / post-compensation (PC) at transmitter (Tx) /receiver (Rx), respectively. In PB analysis, the NL distortion is modelled as an additive term to the signal (A-model), which was developed by improved models such as the additive-multiplicative (AM) model [1], multiplicative-additive (MA) model [2], and the second-order PB model [3]. Conventionally, PC requires symbol decisions for NL distortion reconstruction, which increases complexity and can lead to error propagation due to the error symbols. In [4], it was proposed to re-map the post-FEC bits to symbols for NL estimation to avoid the error symbol issue, but it suffers from even higher complexity.

In [5], we show that the AM-model approximates the NL fiber channel better than the A-model and MA-model, while PC conducted in the manner of negative AM-model is equivalent to back propagation in the AM-model channel, and a raw Rx signal can directly serve as the input for the PC module, without the need for feedback of symbol decision for NL estimation. We note that [6,7] contain related ideas, though these were presented only briefly and not explored in depth. The analysis in [5] is conducted in an ideal system without laser phase noise (PN). In this paper, we included the influence of the PN at Tx and Rx lasers, and compared the performance of PD and PC joint with conventional carrier phase recovery (CPR) and equalization-enhanced phase noise (EEPN)-free CPR, showing that the algorithm flow that sequentially reverses the channel impairments leads to the best performance.

## Principle of Rx-PC and Dx-PC

Single-polarization notation is used in the following for simplicity, while the simulation is conducted in dual polarization (DP). The signal propagation is described by nonlinear Schrodinger equation $iq_z + \beta_2/2q_{tt} + f(\mathrm{z})\gamma|q|^2 q = 0$, where $q(z,t)$ is the complex envelope of the signal, $\beta_2$ and $\gamma$ are the chromatic dispersion and the nonlinearity coefficient, and $f(z)$ denotes the signal power profile. The Tx signal, $q(0,t) = \sum_k \sqrt{P} s_k u_k(0,t)$, where $s_k$ denotes Tx symbols, $u_k(z,t) \triangleq u(z,t-kT)$, $T$ is the symbol period and $P$ is the signal power. Based on the 1st-order PB, after matched filtering, the NL-distorted $k^{\text{th}}$ Rx symbol is given by $\sqrt{P}s_k + \sqrt{P^3}(\Delta s_k^{\text{P}} + \Delta s_k^{\text{C}})$, where $\Delta s_k^{\text{P}}$,$\Delta s_k^{\text{C}}$ is given by $\sum_{m,n} s_{m+k} s_{n+k} s_{m+n+k}^* C_{m,n}$ with $mn = 0$, $mn \neq 0$, respectively, corresponding to the tangent and circular NL distortion,. $C_{m,n} = i\gamma T \int_0^L f(z) \int_{-\infty}^{+\infty} u_0^*(z,t) u_m(z,t) u_n(z,t) u_{m+n}^*(z,t) dt$. When signal power goes high, the 1st-order PB diverges in signal power and loses accuracy. AM-model was proposed to handle the power divergence issue and includes higher-order NL distortion by heuristically forcing the “tangent” $\Delta s_k^{\text{P}}$ into a “phasal” one, and decorate the whole “circular” component, given by $[\sqrt{P}s_k + \sqrt{P^3}\Delta s_k^{\text{C}}]e^{P\Delta s_k^{\text{P}}/s_k}$.

Given a Tx sequence $\boldsymbol{s} = [s_1, s_2, \ldots, s_N]$, the Rx sequence by the AM-model (with the mean NL phase noise removed) is given by $\boldsymbol{r} \approx [\boldsymbol{s} + \boldsymbol{B_s}]e^{i\widetilde{\boldsymbol{A}}_{\boldsymbol{s}}}$, where $\widetilde{\boldsymbol{A}}_{\boldsymbol{s}} = \boldsymbol{A_s} - \mathbb{E}[\boldsymbol{A_s}]$ is the mean-removed NL phase sequence, and $\boldsymbol{B_s}$ is the circular symmetric NL distortion sequence. The multiplication here is elementwise, and the subscript “$\boldsymbol{s}$” indicates that $\widetilde{\boldsymbol{A}}$, $\boldsymbol{B}$ are calculated based on $\boldsymbol{s}$. The conventional PC (Dx-PC) uses the symbol decisions to approximate the Tx sequence for NL estimation, which is then subtracted from the Rx sequence, i.e., $\hat{\boldsymbol{r}}_{\boldsymbol{d}} = \boldsymbol{r} - \left[(\boldsymbol{d} + \boldsymbol{B_d})e^{i\widetilde{\boldsymbol{A}}_{\boldsymbol{d}}} - \boldsymbol{d}\right]$, as shown in Fig. 1(c). Clearly, the performance is bounded by its “genie-aided” version that uses Tx symbols to estimate NL distortion, namely, Tx-PC, $\hat{\boldsymbol{r}}_{\boldsymbol{s}} = \boldsymbol{r} - \left[(\boldsymbol{s} + \boldsymbol{B_s})e^{i\widetilde{\boldsymbol{A}}_{\boldsymbol{s}}} - \boldsymbol{s}\right]$. If the symbol error rate is high, the NL reconstruction and hence the effectiveness of PC will be affected. Although re-mapping the post-FEC bits can perfectly recover Tx symbols [4], it suffers from high complexity.

In [5], we demonstrated that PD or PC is essentially a back propagation in a perturbative sense. Hence, a PC scheme does not necessarily require the symbol decisions. Instead, it can directly use the noisy Rx symbols, forming a feedforward PC (Fig. 1(d), (e)). The Rx-PC is given by $\hat{\boldsymbol{r}}_{\boldsymbol{r}} = (\boldsymbol{r} - \boldsymbol{B}_{\boldsymbol{r}})e^{-i\tilde{A}_r}$, where the subscript "$\boldsymbol{r}$" indicates the $\tilde{\boldsymbol{A}}$, $\boldsymbol{B}$ are calculated based on the Rx sequence. The Rx-PC is achieved in the sense of back propagation rather than removing the estimated NL distortion, where $\boldsymbol{r}$ is by no mean any approximation of $\boldsymbol{t}$, but precisely the required input by back propagation. In practice, $\boldsymbol{r}$ for PC can be the signal before or after CPR as long as it has arrived at one sample/symbol stage, while the former will be preferred, as shown in the following simulations.

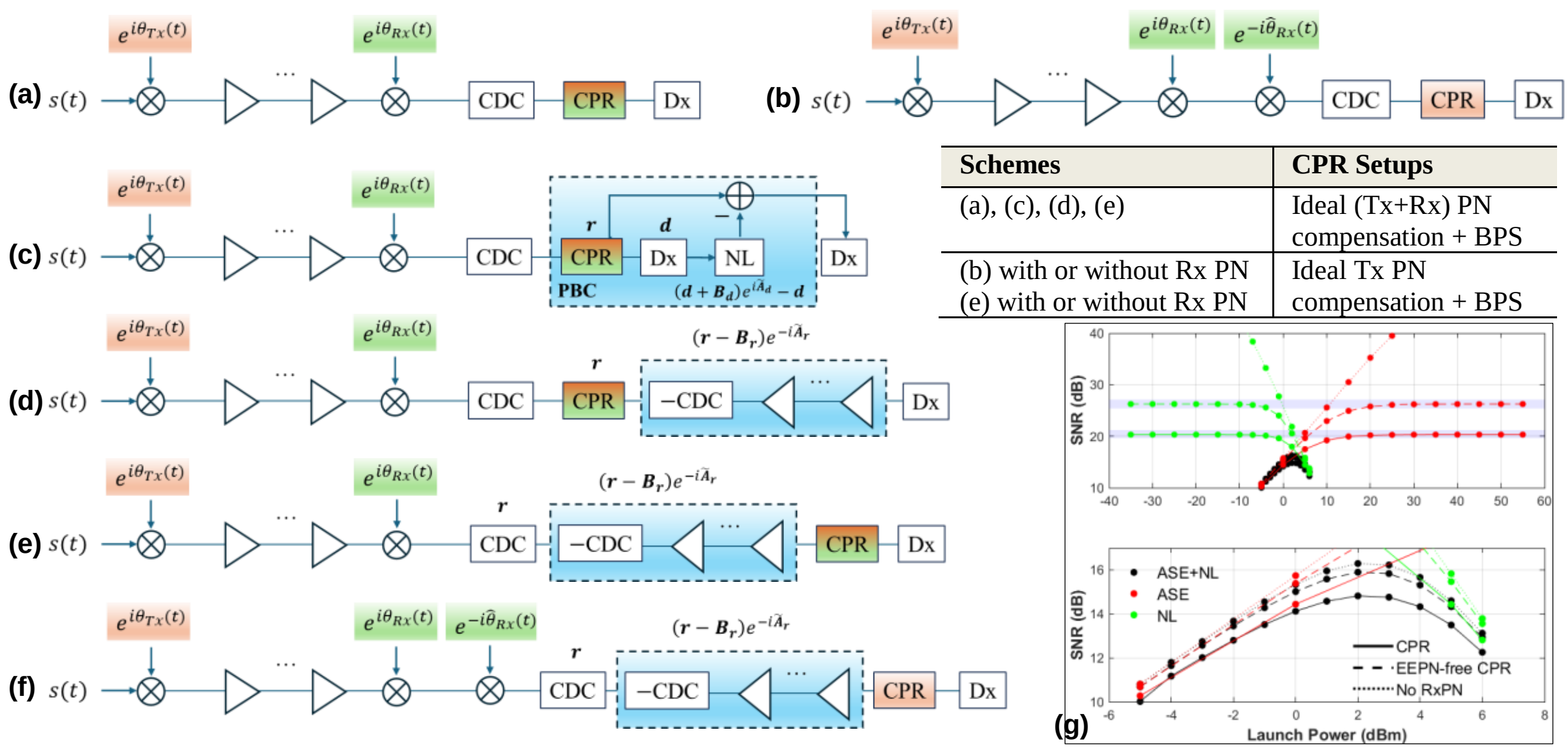

| Schemes | CPR Setups |
|---|---|
| (a), (c), (d), (e) | Ideal (Tx+Rx) PN compensation + BPS |
| (b) with or without Rx PN<br>(e) with or without Rx PN | Ideal Tx PN compensation + BPS |

**Fig. 1: (a)** A typical coherent system without PC; **(b)** A coherent system with EEPN-free CPR, $\hat{\boldsymbol{\theta}}_{\boldsymbol{Rx}}(\boldsymbol{t})$ is the estimation of the true Rx PN; **(c)** Dx-PC; **(d)** Rx-based PBC using $\boldsymbol{r}$ after CPR; **(e)** Rx-PC using $\boldsymbol{r}$ before CPR; **(f)** A coherent system with EEPN-free CPR and Rx-PC using $\boldsymbol{r}$ after CDC before final CPR. The blue box indicates the PBC stage. **(g)** The SNR saturation due to EEPN.

When Tx and Rx laser PN are included, the whole channel and the conventional demodulation processes are illustrated in Fig. 1(a). To effectively compensate for each channel impairment, those effects should be undone in the reverse sequence one by one, due to the non-commutativity of those impairments.

### Combination with the EEPN-free CPR

Typically, the chromatic dispersion compensation (CDC) for a coherent system is implemented in the digital domain after the signal has acquired the Rx PN in the receiver. The Tx PN goes through a net-zero CD while the Rx PN only goes through the CDC stage, resulting in EEPN, which has a variance of $\pi^2|\beta_2|L\Delta\upsilon P/T$ (should time 2/3 if a CPR algorithm is applied [8]) where $\Delta\upsilon$ is the 3dB linewidth of the Rx laser. Recently, it was proposed to use pilot tones [9] or residual carrier [10] located at different frequencies to tackle the EEPN issue. The frequency difference between pilot tones/residual carrier leads to a delay of the Tx PN while they share the same Rx PN. The Tx PN can then be well reconstructed by cumulatively summing the difference between the overall PN of the two pilot tones, and hence, Rx PN can then be isolated from the mixture of Tx and Rx PN, and mitigated individually before the CDC stage, thus alleviating the problem of EEPN. With fiber nonlinearity ignored, such a scheme forms an exact reverse link to the transmission link (Fig. 1(b)), hence improving the performance.

With the philosophy of compensation by forming a reverse link, and the fact that Rx-PC is essentially a back propagation. These two schemes can be combined, as shown in Fig. 1(f), forming an exact reverse of the transmission link with fiber nonlinearity included, which is expected to further improve the performance. Note that the full symmetrical link can only be achieved with PC, because it is impossible to incorporate the Tx PN into the PD calculation, which is conducted on the clean symbols before carrier modulation.

### Simulation Results

We conduct single-channel SSFM simulations by transmitting 32768 45 GBaud (DP)-16QAM symbols modulated with a root-raised-cosine waveform (roll-off of 0.05) and an up-sampling rate of 8 samples/symbol. The transmission link consists of $15 \times 100$ km G.652 fiber ($\alpha = 0.2$ dB/km, $\beta_2 = -20.5$ ps$^2$/km, $\gamma = 1.3$/W/km) with EDFA amplification ($\mathrm{NF} = 5$ dB). No polarization-mode dispersion is added, and ideal polarization demultiplexing is assumed. Tx and Rx PN (linewidth from 100 to 1300 kHz) are included, and ideal phase recovery is used for stably compensating the laser PN, while a Blind Phase Searching algorithm is followed (a CPR is inevitable in a practical system), which can further improve the SNR by mitigating the long-correlated components of the residual laser PN and the NL PN. Tx PN is estimated by differentiating the phase curve of two pilot tones separated by 50 GHz, or that of a residual carrier and a 25 GHz pilot tone, which enables EEPN-free CPR. Frequency offset is ignored as it can always be removed by locating the pilot tone/residual carrier spikes in the electrical spectrum. After CDC, the signal is matched-filtered and is down-sampled to 1SpS, followed by various PC schemes, and the corresponding CPR setups are inserted in Fig. 1.

With ideal Tx and Rx, there are only three sources of noise in the system: Amplified Spontaneous Emission (ASE), EEPN, and NL distortion. Fig. 1(g) shows the SNR curve of conventional CPR, EEPN-free CPR, and No Rx PN situation with ASE and/or NL distortion (Tx, Rx linewidth =1000 kHz). As EEPN is proportional to signal power, it leads to SNR saturation in the ASE/NL-only cases, reaching 20.3 dB, consistent with the theoretical prediction of $T/(\frac{2}{3}\pi^2|\beta_2|L\Delta v)$. The SNR is improved by about 6 dB by EEPN-free CPR, translated into a 1.1 dB gain at the optimal launch power when both ASE and NL exist. The case of no Rx PN is included for reference.

Fig. 2 compares the SNR and Q-factor (in dB) of different PC schemes at Tx/Rx linewidths of 200 and 1200 kHz. For Rx-PC, we can either use CDC symbols (which carry the Tx+Rx PN and NL PN), followed by CPR, or perform CPR first to obtain relatively clean symbols, followed by. These two methods are denoted by $\boldsymbol{CDC_{CDC}}$ (Fig. 1(e)) and $\boldsymbol{CPR_{CPR}}$ (Fig. 1(d)), respectively. It is worth noting that the $\boldsymbol{CDC_{CDC}}$ outperforms $\boldsymbol{CPR_{CPR}}$. The CPR will mitigate most of the Tx PN, Rx PN, and partially the NL PN. The Tx PN and NL PN come from the forward transmission link and should be kept for Rx-PC; then their effect in the forward transmission can be effectively reversed by Rx-PC due to its “back propagation” nature. On the other hand, Rx PN ($\theta^r$) harms the PC process as $\sum_{mn} e^{i(\theta^r_{m+k}+\theta^r_{n+k}-\theta^r_{m+n+k})}(r_{m+k}r_{n+k}r^*_{m+n+k})C_{mn}$ , Although $\theta^r$ is roughly slow-varying within the range of $C_{mn}$, i.e., $\theta^r_{m+k}+\theta^r_{n+k}-\theta^r_{m+n+k}\approx\theta^r_k+\delta\theta_k(m,n)$, where $\delta\theta_k(m,n)$ is the small phase error, thus the PC process is roughly transparent to the Rx PN and can be mitigated after PC, the PC process is “smeared” by summing over phase error $\delta\theta_k(m,n)$. Hence, the conventional CPR process has both positive and negative effects on PC, and overall, a negative effect.

The performance of Dx-PC is close to and bounded by the Tx-based PC. In the high-power region, the symbol error results in an inaccurate estimation of the NL distortion and hence worse performance. Although inaccurate estimation also happens at the low-power region, the NL distortion itself is negligible in this region, and thus does not affect the performance much. The performance of both Dx-PC and Tx-based PC is similar to that of Rx-PC in $\boldsymbol{CPR_{CPR}}$ version, worse than the proposed $\boldsymbol{CDC_{CDC}}$ version.

With Tx Rx linewidth of 1200 kHz, pilot separation of 25GHz, by EEPN-free CPR, the peak SNR is improved by 1.3 dB with CDC-only, while the improvement increases to 1.75 dB with PC by $\boldsymbol{CDC_{CDC}}$.This is because mitigating Rx PN not only benefits the CDC process as it avoids the EEPN issue, but also benefits the PC process for the same philosophy of forming a reverse link. The gains reduce to only 0.25 and 0.37 dB, respective for 200 kHz, due to the insignificant asymmetry brought by Rx PN.

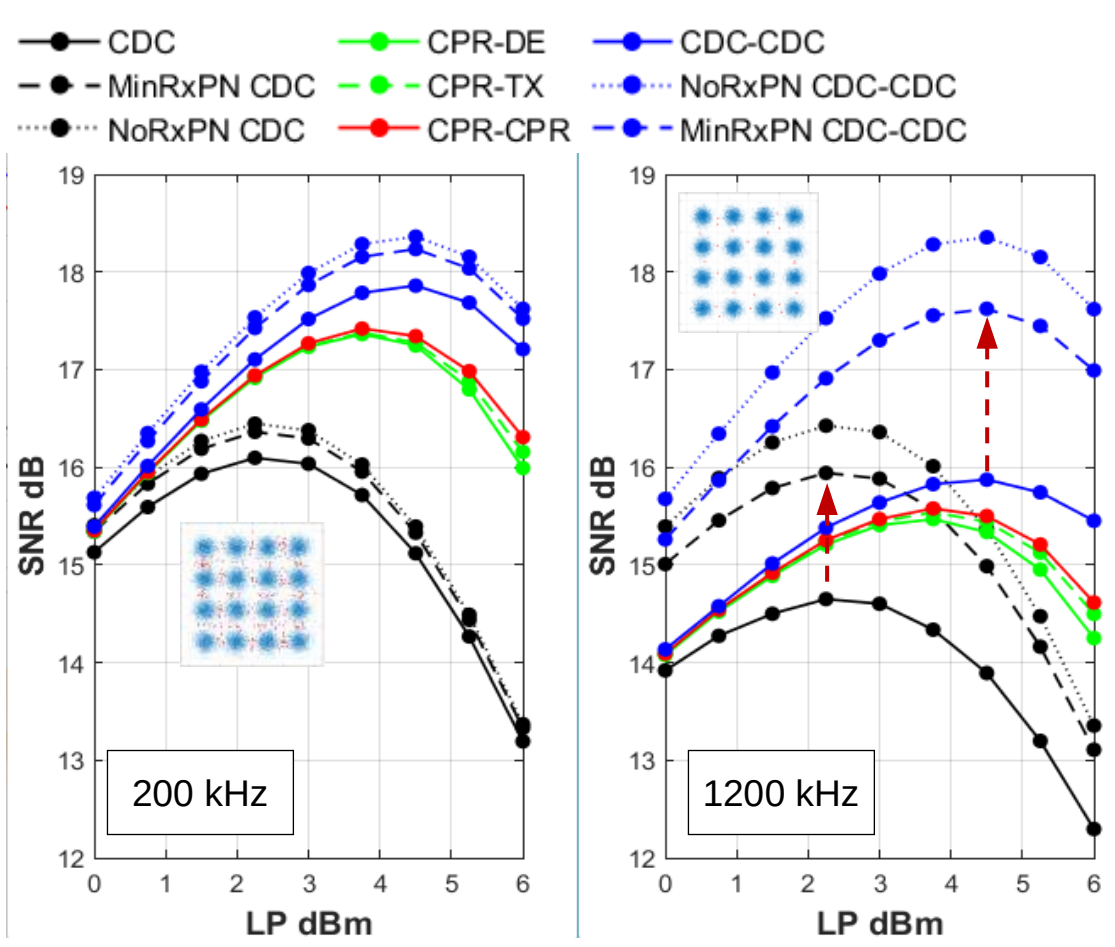

**Fig. 2:** Different PC schemes with Tx Rx linewidth of 200 and 1200kHz. Pilot separation of 25 GHz for EEPN-free CPR.

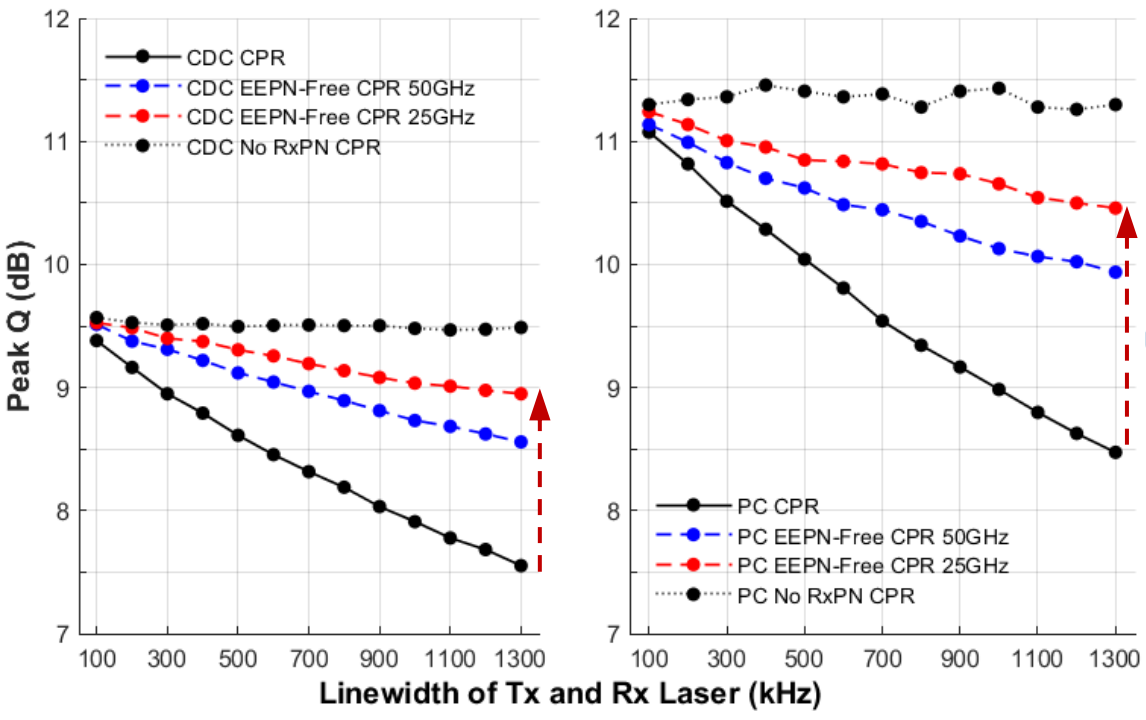

**Fig. 3:** Q-factor (dB) by EEPN-free CPR with and without PC.

We sweep the linewidth from 100 to 1300 kHz and compare the cases of tone-separation of 25 and 50 GHz. At 1300 kHz, which reaches the linewidth of DFB lasers, the Q-factor gain increases to 1.4 dB for CDC and 2 dB for CDC+PC with 25 GHz tone-separation, reducing by ~0.5dB with 50 GHz separation. In the case of no Rx PN (or a perfect compensation of Rx PN), the gain goes to 2 and 2.8 dB, respectively.

## Conclusions

We have demonstrated a feed-forward perturbation-based method that effectively compensates for intra-channel nonlinearities using noisy received signals, which outperforms conventional decision-based schemes. Integrating this method with EEPN-free CPR yields a fully symmetrical link structure, providing significant performance gains.